\documentclass[twocolumn,aps,prb,showpacs]{revtex4}

\usepackage{latexsym,amssymb,float}
\usepackage{setspace}
\usepackage{graphicx}
\usepackage{epsfig}

\newcommand{\e}{\varepsilon}

\newcommand{\f}{\frac}

\begin{document}

\title{Dynamical spin susceptibility and the resonance peak in the
pseudogap region of the underdoped cuprate superconductors}
\author{J.-P. Ismer$^{1,2}$, I. Eremin$^{2,3}$, Dirk K. Morr$^{1,4}$}

\affiliation{$^1$ Institut f\"ur Theoretische Physik, Freie
Universit\"{a}t Berlin, D-14195, Berlin, Germany \\ $^2$ Max-Planck
Institut f\"ur Physik komplexer Systeme, D-01187 Dresden, Germany \\
$^3$ Institut f\"ur Mathematische/Theoretische Physik, Technische
Universit\"at Carolo-Wilhelmina zu Braunschweig, 38106
Braunschweig, Germany \\
$^4$ Department of Physics, University of Illinois at Chicago,
Chicago, IL 60607}
\date{\today}

\begin{abstract}
We present a study of the dynamical spin susceptibility in the
pseudogap region of the high-T$_c$ cuprate superconductors. We
analyze and compare the formation of the so-called resonance peak,
in three different ordered states:  the $d_{x^2-y^2}$-wave
superconducting (DSC) phase, the $d$-density wave (DDW) state, and a
phase with coexisting DDW and DSC order. An analysis of the
resonance's frequency and momentum dependence in all three states
reveals significant differences between them. In particular, in the
DDW state, we find that a nearly dispersionless resonance excitation
exists only in a narrow region around ${\bf Q}=(\pi,\pi)$. At the same time,
in the coexisting DDW and DSC state, the dispersion of the resonance
peak near ${\bf Q}$ is significantly changed from that in the pure
DSC state. Away from $(\pi,\pi)$, however, we find that the form and
dispersion of the resonance excitation in the coexisting
DDW and DSC state and pure DSC state are quite similar. Our results demonstrate
that a detailed experimental measurement of the resonance's
dispersion allows one to distinguish between the underlying phases - a DDW state, a
DSC state, or a coexisting DDW and DSC state - in which the resonance peak emerges.

\end{abstract}

\pacs{71.10.Ca,74.20.Fg,74.25.Ha,74.72.-h} \maketitle

\section{Introduction} One of the most controversial topics in
the field of high-temperature superconductivity is the origin of the
so-called 'pseudogap' phenomenon observed by various experimental
techniques in the underdoped cuprates (for a review see
Ref.~\cite{timusk} and references therein). A large number of
theoretical scenarios have been proposed to explain the origin of
the pseudogap \cite{theory,laughlin}. Among these is the $d$-density
wave (DDW) scenario \cite{laughlin} which was suggested to explain
some of the salient features of the underdoped cuprates such as the
$d_{x^2-y^2}$-wave symmetry of the pseudogap above T$_c$, the
anomalous behavior of the superfluid density\cite{sudip1} and of the
Hall number\cite{sudip2}, as well as the presence of weak (orbital)
antiferromagnetism \cite{eremin}. The DDW-phase is
characterized by circulating bond currents which alternate in space,
break time-reversal symmetry and result in an orbital
(antiferromagnetically ordered) magnetic moment.

In this article, we investigate the momentum and frequency
dependence of the dynamical spin susceptibility, $\chi ({\bf
q},\omega )$, in the underdoped region of the cuprate
superconductors. In particular, we compare the formation of a
resonant spin excitation (the ``resonance peak") in three different
ordered states: the DDW phase, the $d_{x^2-y^2}$-wave
superconducting (DSC) phase, and a phase with coexisting DDW and DSC
order. The observation of the resonance peak in inelastic neutron
scattering (INS) experiments
\cite{rossat,mook0,keimer0,bourges0,bourges,dogan,Fong97,Hay04,stock1}
is one of the key experimental facts in the phenomenology of the
high-T$_{c}$ cuprates. In the optimally and overdoped cuprates, the
resonance peak appears below T$_{c}$ in the dynamical spin
susceptibility at the antiferromagnetic wave vector ${\bf Q}=(\pi
,\pi )$. In the optimally doped cuprates, the resonance's frequency
is $\omega_{res} \approx 41$ meV \cite{rossat,bourges}, a frequency
which decreases with increasing underdoping \cite{Fong97,dogan}. A
number of theoretical scenarios have been suggested for the
appearance of the resonance peak in a superconducting state with
$d_{x^2-y^2}$-wave symmetry \cite{liu,eremin1,other}. In one of
them, the so-called 'spin exciton' scenario \cite{liu,eremin1}, the
resonance peak is attributed to the formation of a particle-hole
bound state below the spin gap (a spin exciton), which is made
possible by the specific momentum dependence of the
$d_{x^2-y^2}$-wave gap. Within this scenario, the structure of spin
excitations in the superconducting state as a function of momentum
and frequency can be directly related to the topology of the Fermi
surface and the phase of the superconducting order parameter. This
scenario agrees well with the experimental data in the
superconducting state of the optimally and overdoped cuprates. In
the underdoped cuprates the resonance-like peak has also been
observed in the pseudogap region above T$_c$ as well as in the
superconducting state \cite{dogan,Fong97,Hay04,stock1}. In this
article, we address the question whether in the underdoped cuprates,
the resonance peak above T$_c$ emerges from the presence of a DDW
state, as first suggested by Tewari {\it et al.}~\cite{sudip1}, and
below T$_c$ from the coexistence of a DSC and DDW phase. To answer
this question, we develop a spin exciton scenario for the pure
DDW-phase as well as the coexisting DSC and DDW phases. By studying
the detailed momentum and frequency dependence of the resonance peak
in both phases, and by comparing it with that in the pure DSC state,
we identify several characteristic features of the resonance peak
that allow one to distinguish between the underlying phases, in
which the resonance peak emerges.

The remainder of the paper is organized as follows: in Secs.~\ref{secDDW} and \ref{secDDWDSC}
we discuss the form of the resonance peak in the pure DDW phase and the coexisting DDW and DSC phase, respectively,
and compare it with that in the pure DSC state. In Sec.~\ref{secconcl} we
summarize our results and conclusions.

\section{Pure DDW state}
\label{secDDW}
Starting point for our calculations in the pure
DDW-state is the effective mean field Hamiltonian
\begin{eqnarray}
H_{DDW} = \sum_{{\bf k},\sigma} \varepsilon_{\bf k}
c^{\dagger}_{{\bf k}, \sigma}c_{{\bf k}, \sigma} + \sum_{{\bf
k},\sigma} i W_{\bf k} c^{\dagger}_{{\bf k}, \sigma}c_{{\bf k+Q},
\sigma} \label{hamDDW}
\end{eqnarray}
where ${\bf Q}=(\pi,\pi)$ is the ordering wavevector of the DDW state,
$W_{\bf k} = \frac{W_0}{2}(\cos k_x -
\cos k_y)$ is the DDW order parameter,
\begin{equation} \varepsilon_{\bf k}=
-2t\left( \cos k_x +\cos k_y\right) - 4t'\cos k_x \cos k_y - \mu
\label{NSdisp}
\end{equation}
is the normal state tight-binding energy dispersion with
$t,t^\prime$ being the hopping elements between nearest and
next-nearest neighbors, respectively, and $\mu$ is the chemical
potential. In the following we use $t=250$meV, $t'/t=-0.4$ and $\mu =
- 1.083t$. The Fermi surface (FS) obtained from Eq.~(\ref{NSdisp})
describes well the FS measured by photoemission experiments on
Bi$_2$Sr$_2$CaCu$_2$O$_{8+\delta}$~\cite{ARPES}. In order to
directly compare the dynamical spin susceptibility in the DDW state
with that in the DSC state, we take the DDW order parameter,
$W_0=42$meV, to be equal to that in the DSC state \cite{eremin1}. We
note here, that the above Hamiltonian can be obtained from a
microscopic Hamiltonian with short-range repulsion or superexchange
interactions \cite{zeyher,dora}.
%
% Fig.1
%
\begin{figure}[h]
\epsfig{file=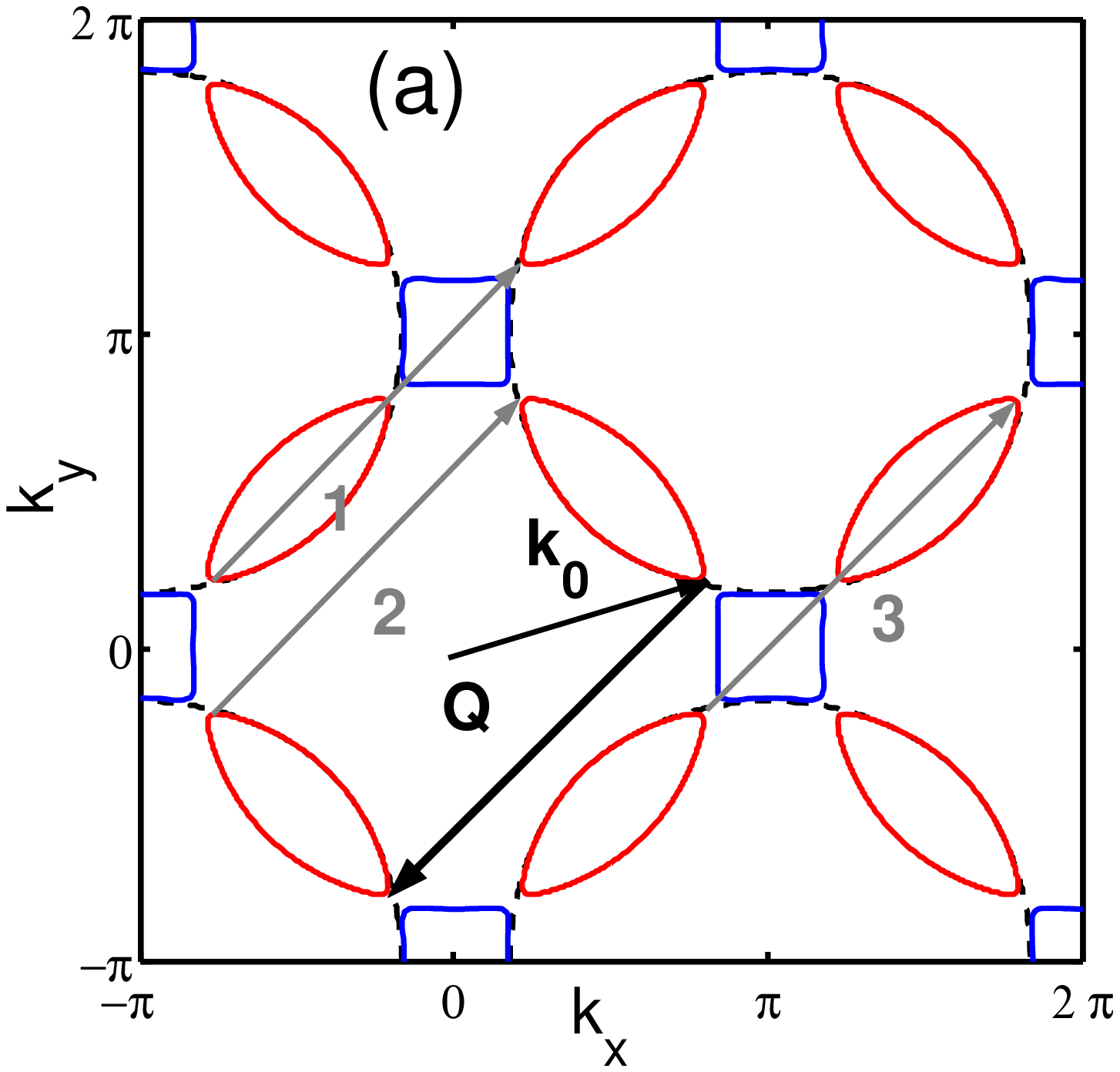,width=6.5cm}
\epsfig{file=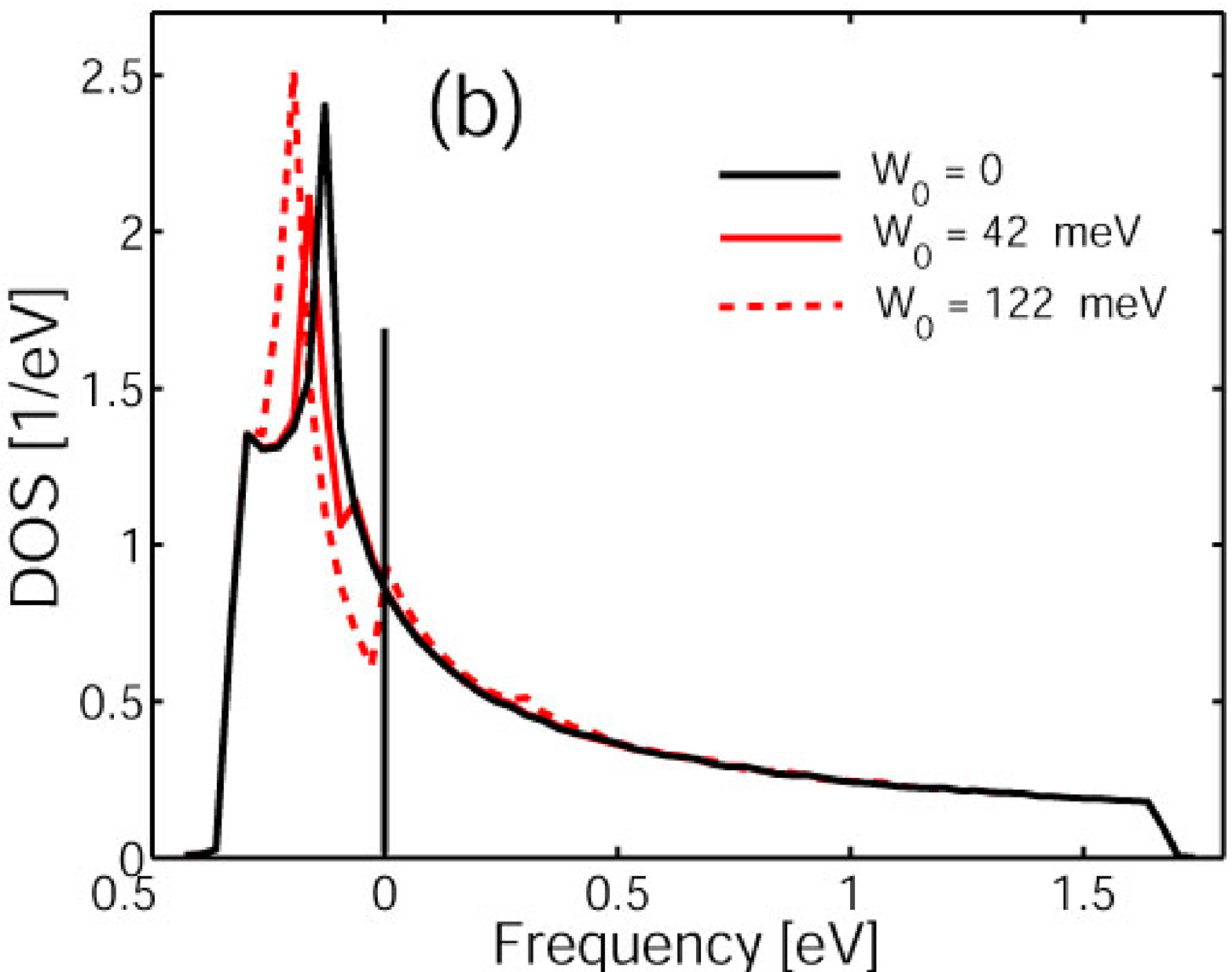,width=6.5cm} \caption{(color online) (a)
Fermi surface in the DDW state for $t=250$meV, $t'/t=-0.4$, $\mu = -
1.083t$ and $W_0=42$meV. The black and gray arrows represent the
magnetic scattering vectors for ${\bf Q}=(\pi,\pi)$ and {\bf Q}$_i =
0.98(\pi,\pi)$, respectively. (b) Calculated DOS for several values
of the DDW gap.} \label{fig1}
\end{figure}
After diagonalizing the Hamiltonian, Eq.(\ref{hamDDW}), one finds
that the excitation spectrum possesses two bands with energy
dispersion
\begin{equation} E^{\pm}_{\bf
k} = \varepsilon^+_{\bf k} \pm \sqrt{\left(\varepsilon^-_{\bf k}
\right)^2+W_{\bf k}^2}\ ,
\end{equation}
where $\varepsilon^{\pm}_{\bf k}=(\varepsilon_{\bf k} \pm
\varepsilon_{\bf k+Q})/2$. In Fig.\ref{fig1}(a) we present the
resulting Fermi surface in the DDW phase. Due to the doubling of the
unit cell in the DDW state, the Fermi surface consists of hole
pockets centered around $(\pm \pi/2, \pm \pi/2)$ and electron
pockets around $(\pm \pi, 0$) and ($0, \pm \pi$). This type of
Fermi surface has not yet been observed experimentally in the
underdoped cuprates, possibly, as has recently been argued, due to
additional interactions between quasiparticles \cite{sudiparp}. For
the above band parameters, the chemical potential lies within both
branches of the excitation spectrum thus preventing the formation of
a gap at the Fermi level. This is clearly visible from
Fig.\ref{fig1}(b) where we plot the density of states (DOS) for
various values of the DDW gap. In particular, one finds that for
$|t'|>W_0/4$ a suppression of (i.e., dip in) the DOS is formed away
from the Fermi level which increases with increasing $W_0$. In
contrast, for $|t'|<W_0/4$, this suppression, which we identify with
the pseudogap, opens at the Fermi level as was noted
previously\cite{carbotte}.  Note that for $t'=0$, the DOS vanishes
at the Fermi level, and the DOS resembles that of a
$d_{x^2-y^2}$-wave superconductor.

In order to compute the dynamical spin susceptibility in the DDW
state, we first introduce the spinor
\begin{equation}
\Psi^\dagger_{{\bf k},\sigma}=\left(c^\dagger_{{\bf
k},\sigma},c^\dagger_{{\bf k+Q},\sigma} \right) \ ,
\end{equation}
where $\sigma$ is the spin index, and the electronic Greens function
in the DDW state is defined as $\hat{G}_{\sigma}({\bf k},
\tau-\tau^\prime) =- \langle {\cal T} \Psi_{{\bf k},\sigma}(\tau)
\Psi^\dagger_{{\bf k},\sigma}(\tau^\prime) \rangle$. The bare
(non-interacting) part of the dynamical spin susceptibility per spin degree of freedom is then given by
\begin{eqnarray}
\chi_0({\bf q}, i \Omega_m) &=& -\frac{T}{8} {\sum_{{\bf k}, n}}^\prime
\mbox{Tr} \left[ {\hat G}({\bf k},i\omega_{n}) \right.
\nonumber \\
& & \left. \times {\hat G}({\bf k+q},i\omega_{n}-i\Omega_{m})
\right] \label{bare}
\end{eqnarray}
where ${\hat G}({\bf k},i\omega_{n})=\hat{G}_{\sigma}({\bf
k},i\omega_{n}) {\hat \sigma}_0$ is the Green's function matrix in
momentum and Matsubara space\cite{mahan}, and the primed sum runs
over  the reduced fermionic Brillouin zone of the DDW state. Note
that with the above definition, one has $\chi_0^{zz}=2 \chi_0$.
After performing the summation over the internal Matsubara
frequencies and analytic continuation to the real frequency axis,
one obtains for the retarded spin susceptibility in the DDW phase
\begin{widetext}
\begin{eqnarray}
\chi_0({\bf q},\omega) &=& \frac{1}{8} {\sum_{\bf k}}^\prime \left(
1+\frac{\varepsilon^-_{\bf k} \varepsilon^-_{\bf k+q}+ W_{\bf k}
W_{\bf k+q}} {\sqrt{ \left( \varepsilon^-_{\bf k} \right)^{2}+
W_{\bf k}^{2}} \sqrt{\left(\varepsilon^-_{\bf k+q} \right)^{2}+
W_{\bf k+q}^2}}  \right) \left( \frac{f(E^+_{\bf k+q})-f(E^+_{\bf
k})}{\omega +i0^+ -E^+_{\bf k+q}+E^+_{\bf k}}+ \frac{f(E^-_{\bf
k+q})-f(E^-_{\bf k})}{\omega+i0^+ -E^-_{\bf k+q}+ E^-_{\bf
k}}\right) \nonumber \\
&& \hspace{-1.5cm} + \left(1-\frac{\varepsilon^-_{\bf k}
\varepsilon^-_{\bf k+q}+W_{\bf k}W_{\bf k+q}}
{\sqrt{\left(\varepsilon^-_{\bf k}\right)^{2}+ W_{\bf k}^{2}}
\sqrt{\left(\varepsilon^-_{\bf k+q}\right)^{2}+ W_{\bf
k+q}^{2}}}\right) \left(\frac{f(E^-_{\bf k+q})-f(E^+_{\bf
k})}{\omega+i0^+-E^-_{\bf k+q} +E^+_{\bf k}}+ \frac{f(E^+_{\bf
k+q})-f(E^-_{\bf k})}{\omega +i0^+ -E^+_{\bf k+q}+ E^-_{\bf
k}}\right) \label{chi_0}
\end{eqnarray}
\end{widetext}
where $f(\epsilon)$ is the Fermi function.

\begin{figure}[h]
\epsfig{file=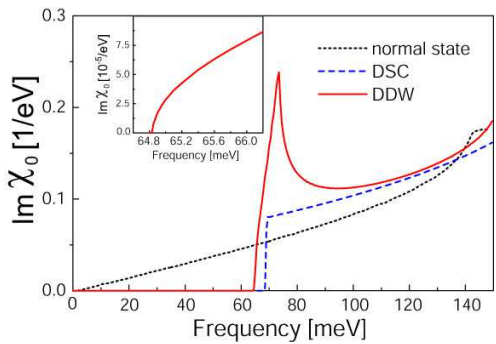,width=6.5cm} \caption{(color online)
Imaginary part of $\chi_0({\bf Q}, \omega)$ at ${\bf Q}=(\pi,\pi)$
in the normal, DSC, and DDW state as a function of frequency. The
momentum dependence of the superconducting gap is taken to be
$\Delta(k)=\Delta_0(\cos{k_x}-\cos{k_y})/2$, with $\Delta_0 =W_0= 42
{\rm meV}$. Inset: Im$\chi_0({\bf Q}, \omega)$ in the DDW phase as a
function of frequency around $\Omega^{DDW}_{cr}$.} \label{fig2}
\end{figure}

We first analyze the behavior of the imaginary part of $\chi_0$ at
${\bf Q}=(\pi,\pi)$, and present in Fig.~\ref{fig2} Im$\chi_0({\bf
Q},\omega)$ as a function of frequency in the normal state, the DSC
state, and the DDW state \cite{comment2} (for the form of $\chi_0$
in the DSC state, see Ref.~\cite{eremin1}).
%
% Fig. 2
%
The behavior of Im$\chi_0({\bf Q},\omega)$ in the normal and DSC
state have been extensively discussed in the literature (see, for
example, Ref.\cite{liu,eremin1,other}). In the normal state
Im$\chi_0$ increases linearly at low frequencies with a slope
determined by the Landau damping rate, while, at higher energies its
behavior is determined by the presence of the van Hove singularity.
In contrast, in the superconducting state the susceptibility is
gapped up to an energy $\Omega^{DSC}_{cr}={\rm min}_{\bf k}
\left(|\Delta_{\bf k}| + |\Delta_{\bf k+Q}|\right)$ where
$\Delta_{\bf k}$ is the superconducting gap and both ${\bf k}$ and
${\bf k+Q}$ lie on the Fermi surface. Due to the symmetry of the
superconducting gap, one finds $\Delta_{\bf k}=- \Delta_{\bf k+Q}$,
resulting in a discontinuous jump of Im$\chi_0$ at
$\Omega^{DSC}_{cr}$ \cite{liu,eremin1}.

In order to discuss the behavior of Im$\chi_0$ in the DDW-state, we
first note that the expression for $\chi_0$ in Eq.(\ref{chi_0})
contains two terms that describe intraband scattering within the
$E^{\pm}_{\bf k}$-bands, and two terms that represent interband
scattering between the two bands. Since $E^{\pm}_{\bf
k+Q}=E^\pm_{\bf k}$ the intraband scattering terms do not contribute
to Im$\chi_0$ at ${\bf Q}$. Moreover, since $E^-_{\bf k} \leq
E^+_{\bf k}$ the first interband scattering term yields a non-zero
contribution to Im$\chi_0$ only for negative frequencies. Thus, only
the second interband scattering term in Eq.(\ref{chi_0}) contributes
to Im$\chi_0$ at ${\bf Q}$. Since the Fermi surfaces of the two
energy bands, $E^{\pm}_{\bf k}$, cannot be connected by the wave
vector ${\bf Q}$, as is evident from Fig.\ref{fig1}(a), Im$\chi_0$
is gapped at low frequencies up to an energy $\Omega^{DDW}_{cr}
\approx 64.8$ meV. The latter is determined by the minimum value of
$E^+_{\bf k}-E^-_{\bf k}=2\sqrt{\left(\varepsilon^-_{\bf k}
\right)^2+W_{\bf k}^2}$ in the DDW Brillouin zone, a condition that is
set by the $\delta$-function arising from the last term in
Eq.(\ref{chi_0}) for Im$\chi_0$. Note that $\varepsilon^-_{\bf k}
\equiv 0$ along the boundary of the DDW Brillouin zone.  Due to the
requirement $sgn(E^+_{\bf k}) \neq sgn(E^-_{\bf k})$, we find that
$\Omega^{DDW}_{cr}=2|W_{\bf k_0}|$ where ${\bf k_0}$ is the momentum
at which the hole pocket around $(\pi/2,\pi/2)$ is intersected by the
DDW Brillouin zone boundary (see Fig.~\ref{fig1}). Moreover, since
$\varepsilon^-_{\bf k+Q}=-\varepsilon^-_{\bf k}$ and $W_{\bf
k+Q}=-W_{\bf k}$,  the second coherence factor in Eq.(\ref{chi_0})
is identical to $2$ for all momenta. Note that there exist two
important differences in Im$\chi_0$ between the DDW and DSC state.
First, in the DSC state, Im$\chi_0 \not = 0$ requires that the
frequency exceeds $\Omega^{DSC}_{cr}=
{\rm min}_{\bf k} \left(|\Delta_{\bf k}| +|\Delta_{\bf k+Q}|\right)$, a
condition which is set by the $\delta$-function in Im$\chi_0$ (see
Eq.(\ref{chi_0})) and simply reflects energy conservation. In
contrast, in the DDW-state, Im$\chi_0 \not = 0$ requires (a) that $\omega -E^+_{\bf k+Q}+
E^-_{\bf k}=0$ for certain momenta ${\bf k}$, and (b) that for the same momenta $f(E^+_{\bf
k+q})-f(E^-_{\bf k}) \not =0$. We find that there exist momenta for
which (a) is satisfied at frequencies $\omega<\Omega^{DDW}_{cr}$,
but that for the same momenta $f(E^+_{\bf k+q})-f(E^-_{\bf k})=0$ (at $T=0$). In other words,
the critical frequency $\Omega^{DDW}_{cr}$ for the onset of a non-zero Im$\chi_0$ is determined by the
difference in the population of the states that are involved in the
scattering process, and not by energy conservation as in the
superconducting state. This qualitative difference between the DDW and the DSC state
bears important consequences: for $\omega> \Omega^{DDW}_{cr}$, one
has Im$\chi_0 \sim \sqrt{\omega- \Omega^{DDW}_{cr}}$, in
contrast to the discontinuous jump of Im$\chi_0$ at $\Omega^{DSC}_{cr}$
in the DSC state (this result for the DDW state differs from that in Ref.~\cite{sudip1} due to the different Fermi surface topology considered here).  This behavior becomes immediately apparent when one plots Im$\chi_0$ in the DDW-state around $\omega=\Omega^{DDW}_{cr}$, as shown in the
inset of Fig.~\ref{fig2}. Note that the number of momenta which are involved in scattering processes and thus contribute to Im$\chi_0$ rapidly increases for
$\omega>\Omega^{DDW}_{cr}$ due to a steeply increasing density of
states of the function $E^+_{\bf k}-E^-_{\bf
k}=2\sqrt{\left(\varepsilon^-_{\bf k} \right)^2+W_{\bf k}^2}$, which
gives rise to the peak in Im$\chi_0$ at $\omega_p \approx 73$ meV.

%
% Fig. 3
%
\begin{figure}[h]
\epsfig{file=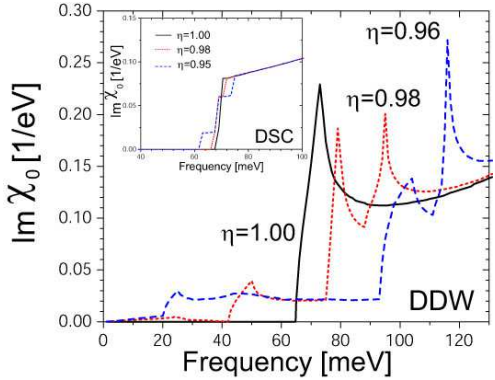,width=6.5cm} \caption{ (color online)
Im$\chi_0$ in the DDW state as a function of frequency for several
momenta ${\bf q}=\eta(\pi,\pi)$ along the diagonal of the magnetic
Brillouin zone (BZ). Inset: Im$\chi_0$ in the DSC state as a
function of frequency for several momenta ${\bf q}=\eta(\pi,\pi)$.}
\label{fig3}
\end{figure}
\begin{figure}[h]
\epsfig{file=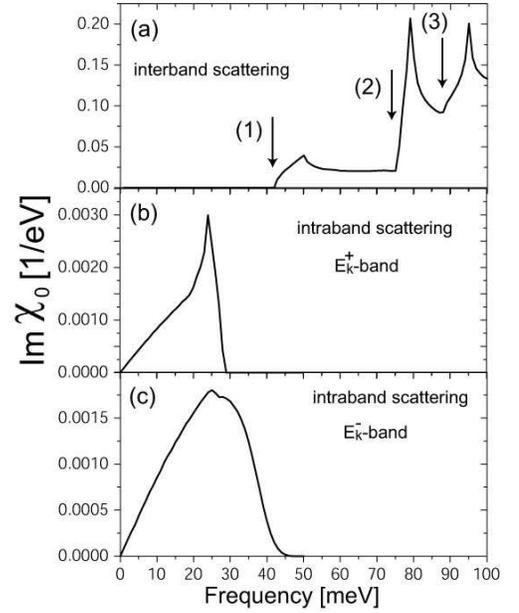,width=6.5cm} \caption{Contributions to
Im$\chi_0$ at ${\bf Q}_i=0.98 {\bf Q}$ from (a) interband
scattering, (b) interband scattering within the $E_k^{+}$-band, and
(c) interband scattering within the $E_k^{-}$-band. The arrows in
(a) indicate the critical frequencies for the opening of interband
scattering channels. The transitions in the fermionic BZ
corresponding to the opening of the interband scattering channels
are shown in Fig.\protect\ref{fig1}(a).} \label{fig_sep}
\end{figure}

For momenta ${\bf q} \not = {\bf Q}$, the behavior of Im$\chi_0$ in
the DDW state is more complex, since in addition to interband
scattering, intraband scattering is now possible. In
Fig.~\ref{fig3}, we plot the frequency dependence of Im$\chi_0$ for
several momenta in the DDW state (for comparison, Im$\chi_0$ in the
DSC state is shown in the inset). We find that as one moves away
from ${\bf Q}=(\pi,\pi)$, several square-root-like increases in
Im$\chi_0$ appear, with the one lowest in energy rapidly decreasing
in frequency. In order to better understand the combined frequency
and momentum dependence of Im$\chi_0$ in the DDW state, we present
in Fig.~\ref{fig_sep} the contributions to Im$\chi_0$ at ${\bf
Q}_i=0.98 {\bf Q}$ from interband scattering [Fig.~\ref{fig_sep}(a)]
and intraband scattering within the $E_k^{+}$-band
[Fig.~\ref{fig_sep}(b)] and $E_k^{-}$-band [Fig.~\ref{fig_sep}(c)]
separately.
%
% Fig. 4
%
Note that while the contribution from intraband scattering is
approximately two orders of magnitude smaller than that from
interband scattering, the former continuously increases from zero
energy, such that Im$\chi_0$ does not any longer exhibit a gap. This
result is valid for all momenta ${\bf q} \not = {\bf Q}$ in the
vicinity of ${\bf Q}$. In contrast, the interband scattering term
possesses three critical frequencies, $\Omega_{cr}^{(i)} (i=1,2,3)$,
which arise from the opening of three non-degenerate scattering
channels that are described by the scattering momenta shown in
Fig.~\ref{fig1}(a). These three critical energies are indicated by
arrows in Fig.~\ref{fig_sep}(a). For all three scattering channels,
the coherence factor is approximately 2. Note that the first and
third scattering channel, which open at $\Omega_{cr}^{(1)} \approx
43$ meV and $\Omega_{cr}^{(3)} \approx 88$ meV and are described by
arrow (1) and (3) in Fig.~\ref{fig1}(a), respectively, connect
momenta ${\bf k}$ and ${\bf k}^\prime$ with ${\bf k}-{\bf
k}^\prime={\bf Q}_i -(\pi,\pi)$ and thus represent umklapp
scattering. In contrast, channel (2) which opens at
$\Omega_{cr}^{(2)} \approx 76$ meV [see arrow (2) in
Fig.~\ref{fig1}(a)] describes direct scattering with ${\bf k}-{\bf
k}^\prime={\bf Q}_i$. The opening of each of these three scattering
channels is accompanied by a square-root like increase of
Im$\chi_0$. Note that the lowest threshold frequency for interband
transitions vanishes at {\bf Q}$_i = 0.94 {\bf Q}$, since this wave
vector connects momenta on the Fermi surfaces of the $E_{\bf k}^{+}$
and $E_{\bf k}^{-}$ bands.

\begin{figure}[h]
\epsfig{file=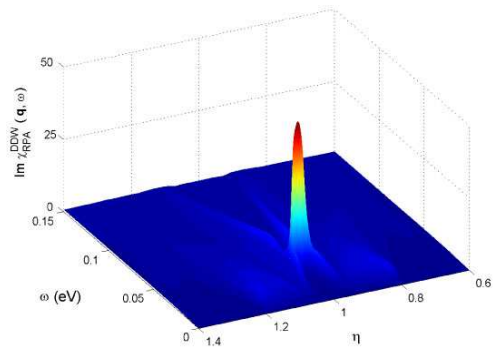,width=6.5cm} \caption{(color online)
Imaginary part of the RPA spin susceptibility in the DDW state as a
function of frequency and momentum along the diagonal of the first
BZ for $U=2.24$eV. } \label{fig5}
\end{figure}

The emergence of a resonance peak in the DDW state can be understood
by considering the dynamical spin susceptibility within the random
phase approximation (RPA). Within this approximation \cite{umklsusc},
the susceptibility (per spin degree of freedom) is given by
\begin{equation}
\chi_{RPA}({\bf q}, \omega) = \frac{ \chi_0({\bf q},
\omega)}{1-U\chi_0({\bf q}, \omega)} \ ,
 \label{susRPA}
\end{equation}
where $U$ is the fermionic four-point vertex. We first consider
${\bf q}={\bf Q}$ and note that in the superconducting state, the
discontinuous jump in Im$\chi_0$ leads to logarithmic singularity in
Re$\chi_0$. As a result, the resonance conditions, $U
\mbox{Re}\chi_0({\bf Q},\omega=\omega_{res})=1$ and
$\mbox{Im}\chi_0({\bf Q},\omega=\omega_{res})=0$, can be fulfilled
simultaneously below the particle-hole continuum for an arbitrarily
small value of $U>0$, leading to the emergence of a resonance peak
as a spin exciton. In contrast, in the DDW state, Im$\chi_0$
exhibits a square-root like frequency dependence above
$\Omega_{cr}^{DDW}$, leading to an increase of Re$\chi_0$ at the
critical frequency, but not to a singularity in Re$\chi_0$.
Specifically, we find
\begin{widetext}
\begin{equation}
{\rm Re} \, \chi_0 = \frac{2 \sqrt{W}}{\pi} \, \alpha \, {\rm Re}
\left[ 2-\sqrt{\frac{\Delta_-}{W}} \arctan{ \left( \sqrt{
\frac{W}{\Delta_-} } \right)} -\sqrt{ \frac{\Delta_+}{W} } \arctan{
\left( \sqrt{\frac{W}{\Delta_+}} \right)} \right] + ...
\end{equation}
\end{widetext}
where $W=E_c-\Omega^{DDW}_{cr}$, $E_c$ is the high energy cut-off
for the square-root like frequency dependence of Im$\chi_0=\alpha
\sqrt{\omega-\Omega^{DDW}_{cr}}$, $\Delta_{\pm}=\Omega^{DDW}_{cr}\pm
\omega$ and the ellipsis denote background contributions to
Re$\chi_0$ that are independent of the opening of a new scattering
channel. The above form of Re$\chi_0$ implies that $U$ now has to
exceed a critical value, $U_c$, before a resonance peak (in the form
of a spin exciton) can emerge in the DDW state. We note, however,
that the values of $U$ typically taken to describe the emergence of
a resonance peak in the DSC state of optimally doped cuprate
superconductors, exceed $U_c$, such that a resonance peak also
emerges in the DDW state. In other words, for $U>U_c$, the resonance
conditions $U \mbox{Re}\chi_0({\bf Q},\omega=\omega_{res})=1$ and
$\mbox{Im}\chi_0({\bf Q},\omega=\omega_{res})=0$ are satisfied in
the DDW state at a frequency $\omega_{res}<\Omega^{DDW}_{cr}$. As a
result, a resonance peak emerges in the RPA spin susceptibility
around ${\bf Q}=(\pi,\pi)$, as shown in Fig.\ref{fig5} (the value of
$U$ is chosen such that in the DSC state, $\omega_{res}=41$ meV).
Away from ${\bf Q}$, the mode becomes rapidly damped due to the
opening of a scattering channel for intraband transitions, as
discussed above. In addition, the lowest critical frequency for
interband transitions rapidly decreases to zero. As a result, the
resonance peak in the DDW state is confined to the immediate
vicinity of ${\bf Q}=(\pi,\pi)$, and shows no significant
dispersion. Note, that the upward and downward structures in
Im$\chi_{RPA}$ visible in Fig.~\ref{fig5} do not represent real
poles in the susceptibility but arise from the frequency structure
of Im$\chi_0$ away from $(\pi,\pi)$. This momentum dependence of the
'spin exciton' in the DDW state stands in stark contrast to the
dispersion of the resonance peak in the DSC state \cite{eremin1}
(see Fig.~\ref{dispcomp}).
%
% Fig. 5
%

\section{Coexisting DDW and DSC phases}
\label{secDDWDSC}

We next consider a state with coexisting DDW and DSC order whose
mean-field Hamiltonian is given by
\begin{eqnarray}
\mathcal{H}^{DSC+DDW} = \sum_{k}\psi_{k}^{\dag}H(k)\psi_{k} \quad,
\label{hddwdsc}
\end{eqnarray}
where $\psi_{\bf k}^{\dag}=\left(c_{{\bf
k}\uparrow}^{\dag},c_{{\bf k+Q}\uparrow}^{\dag},c_{-{\bf
k}\downarrow},c_{{\bf -k-Q}\downarrow}\right)$ and \cite{sudip1}
\begin{eqnarray}
H_{\bf k} &=&\left(\begin{array}{cccc} \varepsilon_{\bf k} & i
W_{\bf k} &
\Delta_{\bf k} & 0 \\
-i W_{\bf k} & \varepsilon_{\bf k+Q} & 0 & -\Delta_{\bf k} \\
\Delta_{\bf k} & 0 & -\varepsilon_{k} & i W_{\bf k} \\
0 & -\Delta_{\bf k} & -iW_{\bf k} & -\varepsilon_{\bf k+Q}
\end{array}\right) \quad.
\label{hk}
\end{eqnarray}
The energy bands arising from diagonalizing the Hamiltonian in
Eq.(\ref{hk}) are given by
\begin{equation}
\Omega^{\pm}_{{\bf k}}   =  \sqrt{\left(E^{\pm}_{\bf k}\right)^2 +
\Delta_{\bf k}^{2}} \ ,
\end{equation}
with $E^{\pm}_{\bf k}$ being the energy bands of the pure DDW state
given above. The bare susceptibility, $\chi_0$ in the coexisting
phase can again be calculated using Eq.(\ref{bare}) with the only
difference that the Green's function $\hat{G}_{\sigma}({\bf
k},i\omega_{n})$ is now a $(4 \times 4)$ matrix. The full expression
for $\chi_0$ in the coexistence phase is somewhat lengthy and
therefore given in Appendix \ref{appendix}.

In Fig.~\ref{Chi0_DDWDSC} we present Im$\chi_0$ as a function of
frequency for several momenta ${\bf q}=\eta(\pi,\pi)$ along the diagonal of the magnetic BZ.
%
% Fig. 6
%
\begin{figure}[h]
\epsfig{file=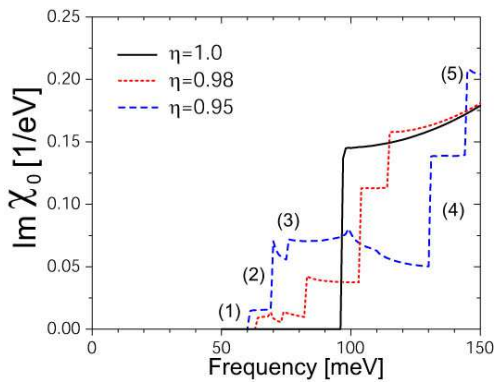,width=6.5cm}
 \caption{(color online) Im $\chi_0$ as a function of frequency in the
 coexisting DDW and DSC state for several momentum ${\bf q}=\eta(\pi,\pi)$ along the diagonal of the magnetic BZ.} \label{Chi0_DDWDSC}
\end{figure}
At ${\bf Q}=(\pi,\pi)$ ($\eta=1.0$), Im$\chi_0$ exhibits a single
discontinuous jumps at the critical frequency, $\Omega^{coex}_{cr}=97$ meV.
The magnetic scattering associated with the opening of this
scattering channel connects the ``hot spots" in the fermionic BZ,
i.e., those momenta ${\bf k}$ and ${\bf k+Q}$ for which
$\varepsilon_{\bf k}=\varepsilon_{\bf k+Q}=0$. Correspondingly, the
critical frequency is given by $\Omega^{coex}_{cr}=2 \sqrt{\Delta^2({\bf
k}_{hs})+W^2({\bf k}_{hs})}$ where ${\bf k}_{hs}$ is the momentum of
the hot spots. In contrast, away from ${\bf Q}=(\pi,\pi)$, we find
that Im$\chi_0$ exhibits 5 discontinuous jumps at critical
frequencies, $\Omega_{cr}^{(i)}$ with $i=1,..,5$ indicating the
opening of new scattering channels (for $\eta=0.95$ these five
discontinuous jumps are labeled in Fig.~\ref{Chi0_DDWDSC}). Note
that in the coexistence
phase, the opening of a new scattering channel is accompanied by a
discontinuous jump, similar to the pure DSC state, but in contrast
to the DDW state, as discussed above. The momentum dependence of
these critical frequencies is shown in Fig.~\ref{jumps}(a).
%
% Fig.7
%
\begin{figure}[h]
\epsfig{file=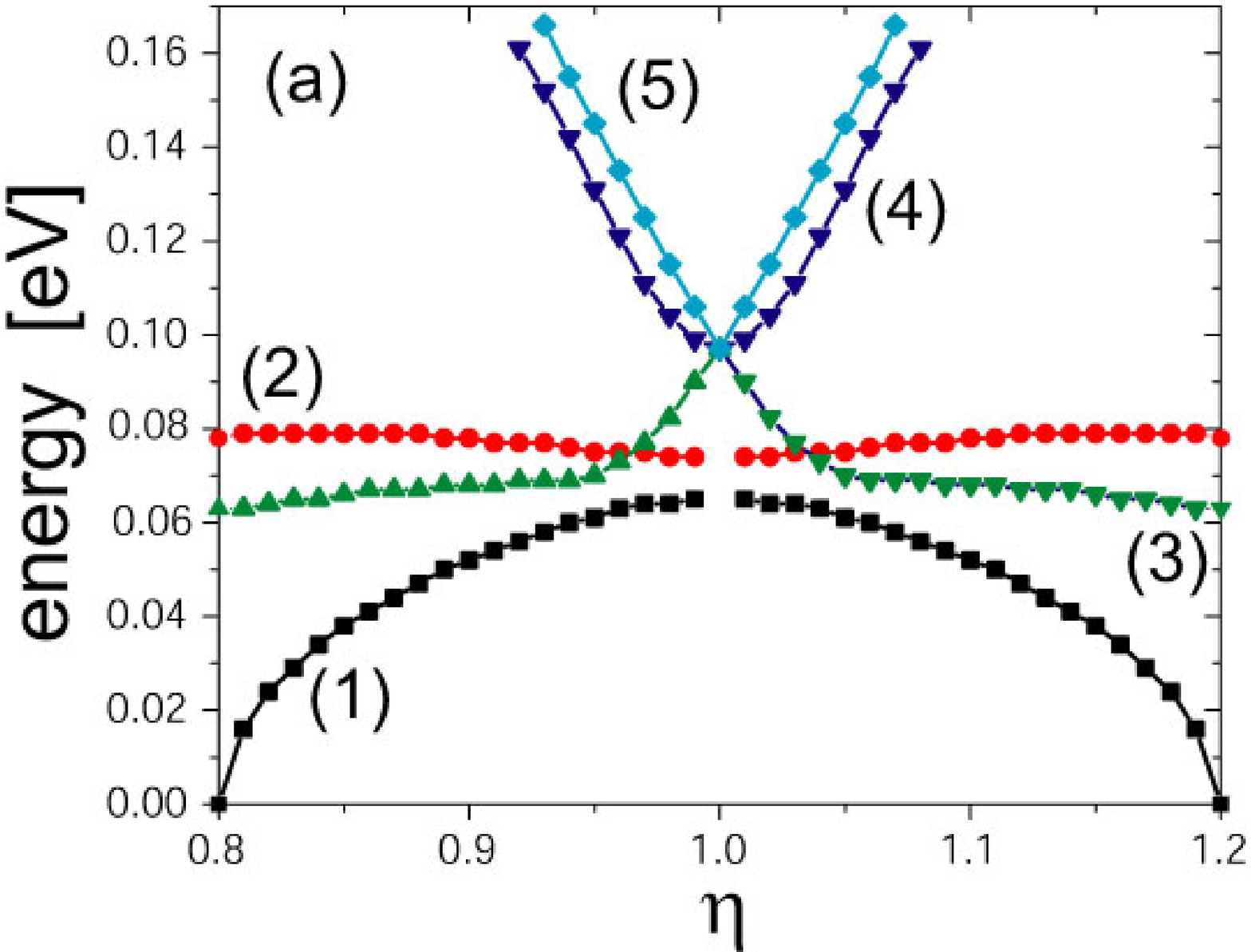,width=6.5cm}
\epsfig{file=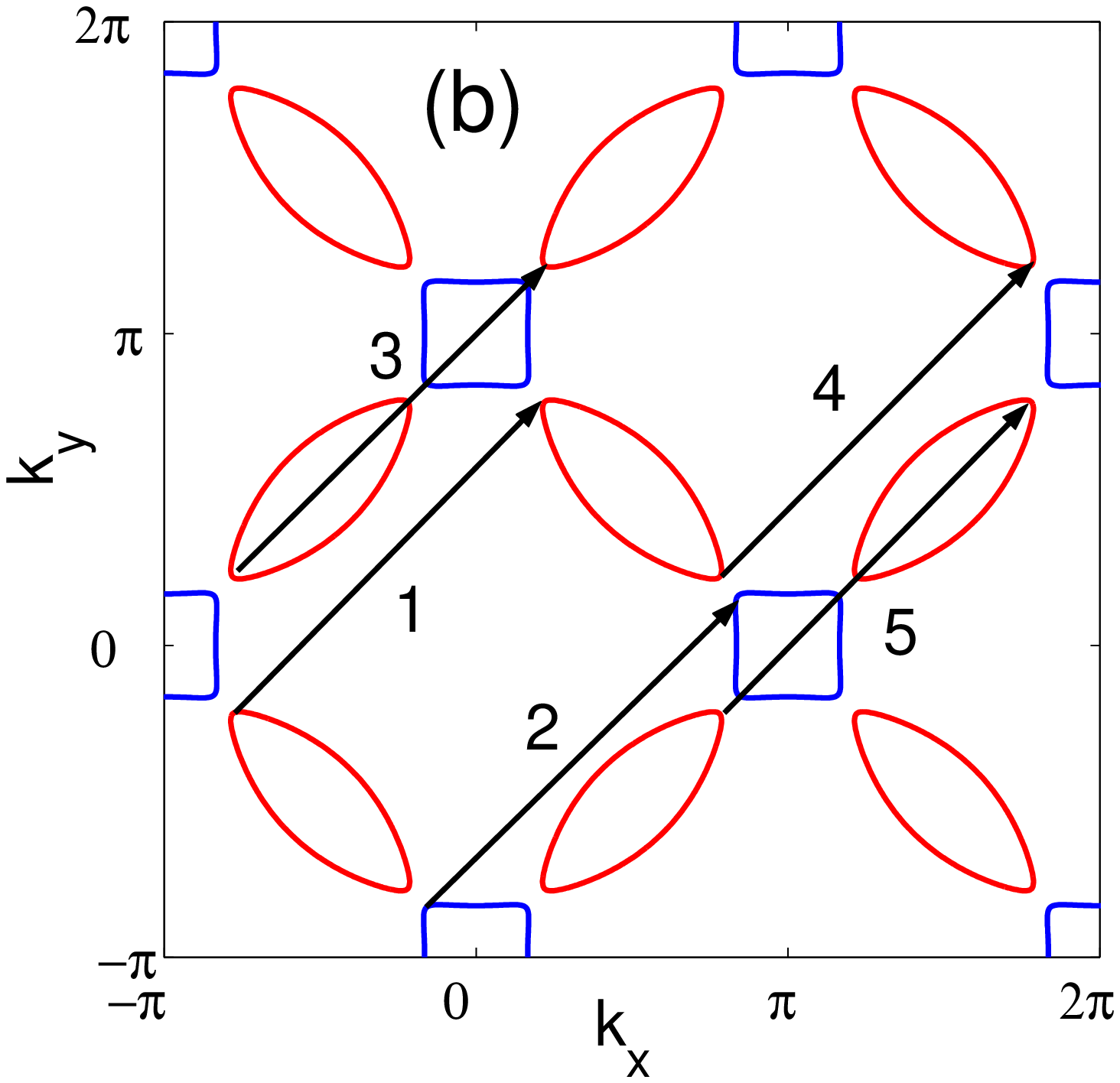,width=6.5cm} \caption{(color online)(a)
Momentum dependence of the critical frequencies $\Omega_{cr}^{(i)}$
($i=1,..,5$). (b) FS of the pure DDW state and magnetic scattering
vectors of the 5 scattering channels that open at
$\Omega_{cr}^{(i)}$.} \label{jumps}
\end{figure}
At $\Omega_{cr}^{(1)}$ [$\Omega_{cr}^{(2)}$], a scattering channel
for intraband scattering within the $\Omega^{+}_{{\bf k}}$
($\Omega^{-}_{{\bf k}}$) band opens, and Im$ \chi_0$ acquires a
non-zero contribution from $\chi^{(2)}$ ($\chi^{(5)}$) given in
Eq.(\ref{chi2}) [Eq.(\ref{chi5})] of Appendix \ref{appendix}. As
follows directly from Eqs.(\ref{chi2}) and (\ref{chi5}), the
coherence factors associated with these two scattering processes
vanish identically at ${\bf Q}=(\pi,\pi)$, and hence, no value for
$\Omega_{cr}^{(1,2)}$ can be defined at this momentum. However, away
from ${\bf Q}=(\pi,\pi)$, the coherence factors are not longer zero,
and two discontinuous jumps appear in Im$\chi_0$ that are associated
with the opening of two new scattering channels at
$\Omega_{cr}^{(1,2)}$. Note that the magnitude of the jumps at
$\Omega_{cr}^{(1,2)}$ increases as one moves away from ${\bf
Q}=(\pi,\pi)$, which is a direct consequence of the increasing
coherence factors. Similar to the superconducting state the lowest
critical frequency, $\Omega_{cr}^{(1)}$, reaches zero at
$\eta=0.8(\pi, \pi)$. In contrast, at $\Omega_{cr}^{(3,4,5)}$ new
scattering channels for interband scattering between the
$\Omega^{+}_{{\bf k}}$ and $\Omega^{-}_{{\bf k}}$ bands are opened.
These three critical frequencies are degenerate at ${\bf
Q}=(\pi,\pi)$, but this degeneracy is lifted for ${\bf q} \not =
{\bf Q}$, as follows immediately from Fig.~\ref{jumps}(a). For
$\eta=0.98$, we present in Fig.~\ref{jumps}(b) the scattering
momenta that are associated with the opening of the above discussed
five scattering channels. For completeness, we also present the
Fermi surfaces for the $E^{\pm}_k$ bands. Note, that the scattering
vectors (1) and (2) describe intraband scattering within the
$\Omega^{+}_{{\bf k}}$ and $\Omega^{-}_{{\bf k}}$ bands, while the
scattering vectors (3), (4) and (5) represent interband scattering.
The scattering vectors (3), (4), and (5) are identical to those
present in the pure DDW state [for comparison, see Fig.1 (a)].

\begin{figure}[h]
\epsfig{file=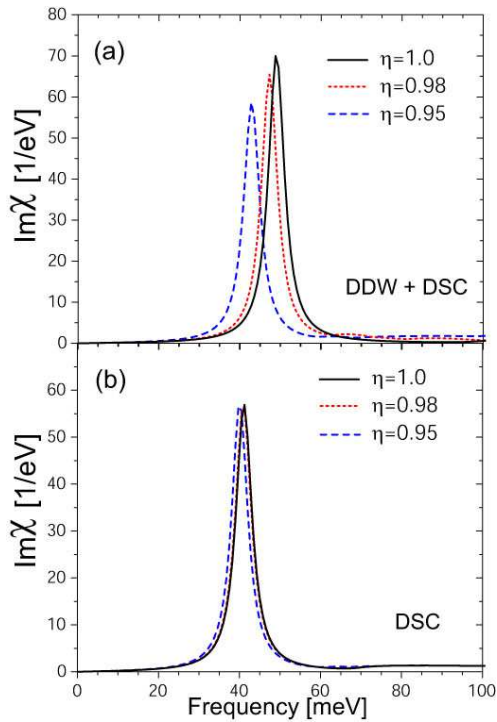,width=6.5cm} \caption{(color online) (a)
RPA spin susceptibility as a function of frequency in the coexisting
DDW and DSC state for several momenta. (b) RPA spin susceptibility
as a function of frequency in the pure DSC state for several
momenta. } \label{RPAchi}
\end{figure}

\begin{figure}[h]
\epsfig{file=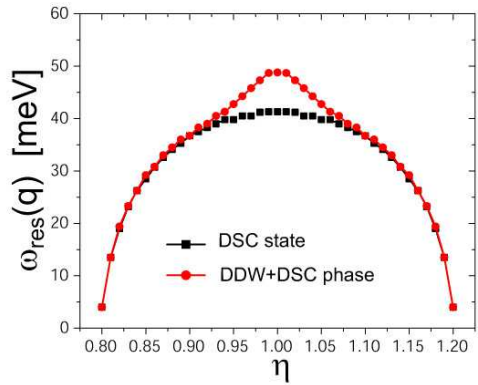,width=6.5cm}
 \caption{(color online) Dispersion of the resonance peak in the coexisting
DDW and DSC state, as well as in the pure DSC state.}
\label{dispcomp}
\end{figure}

In Fig.~\ref{RPAchi} we present the RPA susceptibility in the
coexisting DDW and DSC phase.
%
% Fig.8
%
Similarly to the pure DSC state, the discontinuous jump in
Im$\chi_0$ in the coexistence phase is accompanied by a logarithmic
divergence in Re$\chi_0$, which in turn gives rise to a resonance
peak below the particle-hole continuum for an arbitrary small
fermionic interaction. A comparison with the RPA susceptibility in
the pure DSC state shown in Fig.~\ref{RPAchi}(b) reveals that the
frequency position of the resonances peak varies more quickly with
momentum (near ${\bf Q}=(\pi,\pi)$) in the coexistence phase than in
the pure DSC state. This difference becomes particularly evident
when one plots the dispersion of the resonance peak both in the
coexistence phase and the pure DSC state, as shown in
Fig.~\ref{dispcomp}.
%
% Fig.9
%
We find that the difference in the dispersion of the resonance peaks
is particularly pronounced around ${\bf Q}$ ($0.95 \lesssim \eta
\lesssim 1.05$), with a more cusp-like dispersion in the coexistence
phase. This cusp follows the form of the particle-hole continuum in
the vicinity of ${\bf Q}$ in the coexistence phase, as is evident
from Fig.~\ref{jumps}(a). Thus the dispersion of the resonance peak
directly reflects the different momentum dependence of the
particle-hole continuum in the vicinity of ${\bf Q}$ in the
coexisting DDW and DSC state and the pure DSC state. However, away
from ${\bf Q}$, the particle-hole continuum, as well as the
dispersion of the resonance peak are quite similar in both phases.

\section{Conclusions}
\label{secconcl}

In conclusion, we have analyzed the momentum and frequency
dependence of the dynamical spin susceptibility in the pure DDW
state and the phase with coexisting DDW and DSC order. We find that
due to the opening of a spin gap in Im$\chi_0$ in both phases, a
resonance peak emerges below the particle-hole continuum. However,
in the DDW state, Im$\chi_0$ exhibits a square-root like increase at
the critical frequencies, in contrast to the coexisting DDW and DSC
phases (or the pure DSC phase), where the onset of Im$\chi_0\not =
0$ is accompanied by a discontinuous jump. As a result, Re$\chi_0$
in the DDW state does not exhibit a divergence, but simply an
enhancement at the critical frequency, and hence, a finite fermionic
interaction strength is necessary for the emergence of a resonance
peak in the DDW state. This result is qualitatively different from
the coexisting DDW and DSC phase and the pure DSC state where a
resonance peak emerges for an infinitesimally small interaction
strength. We note, however, that for the strength of the fermionic
interaction usually taken to describe the resonance peak in the DSC
state, a resonance peak also emerges in the DDW state. Moreover, we
find that the resonance peak in the DDW state is basically
dispersionless and confined to the vicinity of ${\bf Q}=(\pi,\pi)$
due to the form of the particle-hole continuum in the DDW state. In
contrast, the dispersion of the resonance peak in the coexisting DDW
and DSC state is similar to that in the pure DSC state, with the
exception that in the vicinity of ${\bf Q}$, the former exhibits a
cusp. These results show that the detailed momentum and frequency
dependence of the resonance peak is different in all three phases,
the pure DDW, pure DSC and coexisting DDW and DSC phases. Thus, a
detailed experimental study of the resonance peak in the underdoped
cuprates permits one to identify the nature of the underlying phase
in which the resonance peaks emerges. We believe, however, that the
currently available experimental INS data do not yet allow an
unambiguous conclusion with regards to the nature of the phases
present in the underdoped cuprate superconductors.\\

D.K.M. acknowledges financial support by the Alexander von Humboldt
Foundation, the National Science Foundation under Grant No.
DMR-0513415 and the U.S. Department of Energy under Award No.
DE-FG02-05ER46225.

\appendix

\begin{widetext}

\section{Dynamical spin susceptibility in the regime of coexisting
DDW+DSC phases} \label{appendix} In the coexisting DDW and $d$-wave
superconducting phase, the susceptibility is given by
\begin{equation}
\chi({\bf q},\omega)=\sum_i \chi^{(i)}({\bf q},\omega)
\end{equation}
where
\begin{eqnarray}
\chi^{(1)}({\bf q},\omega)&=&\f{1}{16} \sum_{\bf k} \left(1+\f{E_{\bf
k}^{+}E_{\bf k+q}^{+}+\Delta_{\bf k}\Delta_{\bf
k+q}}{\Omega^{+}_{{\bf k}}\Omega^{+}_{{\bf k+q}}}\right)
\left(1+\f{\e^-_{\bf k}\e^-_{\bf k+q}+W_{\bf k}W_{\bf k+q}}
{\sqrt{\left(\e^-_{\bf k}\right)^{2}+W_{\bf
k}^{2}}\sqrt{\left(\e^-_{\bf k+q}\right)^{2}+W_{\bf
k+q}^{2}}}\right)
\f{ f(\Omega^{+}_{\bf k+q})-f(\Omega^{+}_{\bf
k})}{\omega-\Omega^{+}_{\bf k+q}+\Omega^{+}_{\bf k}+i\delta}
\nonumber\\
\label{chi1}
\end{eqnarray}
\begin{eqnarray}
\chi^{(2)}({\bf q},\omega) & = & \f{1}{32} \sum_{\bf k}
\left(1-\f{E_{\bf k}^{+}E_{\bf k+q}^{+}+\Delta_{\bf k}\Delta_{\bf
k+q}}{\Omega^+_{\bf k}\Omega^+_{\bf k+q}}\right)
\left(1+\f{\e^-_{\bf k}\e^-_{\bf k+q}+W_{\bf k}W_{\bf k+q}}
{\sqrt{\left(\e^-_{\bf k}\right)^{2}+W_{\bf
k}^{2}}\sqrt{\left(\e^-_{\bf k+q}\right)^{2}+W_{\bf
k+q}^{2}}}\right) \nonumber
\\&& \times \left(\f{1-f(\Omega^+_{\bf k+q})-f(\Omega^+_{\bf k})}{\omega+\Omega^+_{\bf k+q}+
\Omega^+_{ \bf k}+i\delta} +\f{f(\Omega^+_{\bf k+q})+f(\Omega^+_{\bf
k})-1}{\omega-\Omega^+_{\bf k+q}-\Omega^+_{\bf k}+i\delta}\right)
\label{chi2}
\end{eqnarray}
\begin{eqnarray}
\chi^{(3)}({\bf q},\omega)&=&\f{1}{16} \sum_{\bf k} \left(1+\f{E_{\bf
k}^{+}E_{\bf k+q}^{-}+\Delta_{\bf k}\Delta_{\bf k+q}}{\Omega^+_{\bf
k}\Omega^-_{\bf k+q}}\right) \left(1-\f{ \e^-_{\bf k}\e^-_{\bf
k+q}+W_{\bf k}W_{\bf k+q}} {\sqrt{\left(\e^-_{\bf
k}\right)^{2}+W_{\bf k}^{2}}\sqrt{\left(\e^-_{\bf
k+q}\right)^{2}+W_{\bf k+q}^{2}}}\right) \nonumber
\\&& \times \left(\f{f(\Omega^-_{\bf k+q})-f(\Omega^+_{\bf k})}
{\omega-\Omega^-_{\bf k+q}+\Omega^+_{\bf k} +i\delta}
+\f{f(\Omega^+_{\bf k})-f(\Omega^-_{\bf k+q})}{\omega-\Omega^+_{\bf
k}+\Omega^-_{\bf k+q} +i\delta}\right) \label{chi3}
\end{eqnarray}
\begin{eqnarray}
\chi^{(4)}({\bf q},\omega)&=&\f{1}{16} \sum_{\bf k} \left(1-\f{E_{\bf
k}^{+}E_{\bf k+q}^{-}+\Delta_{\bf k}\Delta_{\bf k+q}}{\Omega^+_{\bf
k}\Omega^-_{\bf k+q}}\right) \left(1-\f{\e^-_{\bf k}\e^-_{\bf
k+q}+W_{\bf k}W_{\bf k+q}} {\sqrt{\left(\e^-_{\bf
k}\right)^{2}+W_{\bf k}^{2}}\sqrt{\left(\e^-_{\bf
k+q}\right)^{2}+W_{\bf k+q}^{2}}}\right) \nonumber
\\&& \times \left(\f{1-f(\Omega^-_{\bf k+q})-f(\Omega^+_{\bf k})}
{\omega+\Omega^+_{\bf k}+\Omega^-_{\bf k+q}+i\delta}
+\f{f(\Omega^+_{\bf k})+f(\Omega^-_{\bf
k+q})-1}{\omega-\Omega^-_{\bf k+q}-\Omega^+_{\bf k} +i\delta}\right)
\label{chi4}
\end{eqnarray}
\begin{eqnarray}
\chi^{(5)}({\bf q},\omega)&=&\f{1}{32} \sum_{\bf k} \left(1-\f{E_{\bf
k}^{-}E_{\bf k+q}^{-}+\Delta_{\bf k}\Delta_{\bf k+q}}{\Omega^-_{\bf
k}\Omega^-_{\bf k+q}}\right) \times\left(1+\f{\e^-_{\bf k}\e^-_{\bf
k+q}+W_{\bf k}W_{\bf k+q}} {\sqrt{\left(\e^-_{\bf
k}\right)^{2}+W_{\bf k}^{2}}\sqrt{\left(\e^-_{\bf
k+q}\right)^{2}+W_{\bf k+q}^{2}}}\right) \nonumber
\\&& \times \left(\f{1-f(\Omega^-_{\bf k+q})-f(\Omega^-_{\bf k})}{\omega+\Omega^-_{\bf k}+
\Omega^-_{\bf k+q}+i\delta} +\f{f(\Omega^-_{\bf k})+f(\Omega^-_{\bf
k+q})-1}{\omega-\Omega^-_{\bf k+q}-\Omega^-_{\bf k}+i\delta}\right)
\label{chi5}
\end{eqnarray}
\begin{eqnarray}
\chi^{(6)}({\bf q},\omega)&=&\f{1}{16} \sum_{\bf k} \left(1+\f{E_{\bf
k}^{-}E_{\bf k+q}^{-}+\Delta_{\bf k}\Delta_{\bf k+q}}{\Omega^-_{\bf
k}\Omega^-_{\bf k+q}}\right) \left(1+\f{\e^-_{\bf k}\e^-_{\bf
k+q}+W_{\bf k}W_{\bf k+q}} {\sqrt{\left(\e^-_{\bf
k}\right)^{2}+W_{\bf k}^{2}}\sqrt{\left(\e^-_{\bf
k+q}\right)^{2}+W_{\bf k+q}^{2}}}\right) \f{f(\Omega^-_{\bf
k+q})-f(\Omega^-_{\bf k})}{\omega-\Omega^-_{\bf k+q}+\Omega^-_{\bf
k}+i\delta} \nonumber \\
\label{chi6}
\end{eqnarray}
\end{widetext}


\begin{thebibliography}{99}
\bibitem{comment2} The results for Im$\chi_0$ were obtained by using the tetrahedron
method \cite{taut} to evaluate Eq.(\ref{chi_0}).
\end{thebibliography}
\end{document}